\documentclass[showpacs,aps,amsmath,amssymb,superscriptaddress,twocolumn,prb]{revtex4-1}
\usepackage{bm}
\usepackage{graphicx}
\usepackage[english]{babel}
\newcommand{\B}{\text{\scriptsize res}}
\newcommand{\s}{\text{\scriptsize sys}}
\newcommand{\SB}{\text{\scriptsize sys-res}}
\newcommand{\T}{{\rm total}}

\newcommand{\nl}{\nonumber \\}
\newcommand{\la}{\langle}
\newcommand{\ra}{\rangle}

\newcommand{\ep}{\epsilon}
\newcommand{\w}{\omega}

\newcommand{\Sec}[1]{Sec.\,\ref{#1}}

\newcommand{\be}{\begin{equation}}
\newcommand{\ee}{\end{equation}}
\newcommand{\bea}{\begin{eqnarray}}
\newcommand{\eea}{\end{eqnarray}}
\newcommand{\bsube}{\begin{subequations}}
\newcommand{\esube}{\end{subequations}}
\newcommand{\Eq}[1]{Eq.\,\eqref{#1}}
\newcommand{\Eqs}[1]{Eqs.\,\eqref{#1}}
\newcommand{\Fig}[1]{Fig.\,\ref{#1}}

\newcommand{\bfG}{\bm{\mathcal{G}}}

\newcommand{\comments}[1]{}

\begin{document}

\title{Hierarchical equations of motion for impurity solver in
dynamical mean-field theory}

\author{Dong~Hou}
\affiliation{Hefei National Laboratory for Physical Sciences at the
Microscale, University of Science and Technology of China, Hefei, Anhui
230026, China}
\affiliation{Synergetic Innovation Center of Quantum Information and Quantum Physics,
University of Science and Technology of China, Hefei, Anhui 230026, China}
\affiliation{Department of Chemistry, Hong Kong University of Science
and Technology, Hong Kong, China}

\author{Rulin~Wang}
\affiliation{Department of Physics, University of Science and
Technology of China, Hefei, Anhui 230026, China}

\author{Xiao~Zheng} \email{xz58@ustc.edu.cn}
\affiliation{Hefei National Laboratory for Physical Sciences at the
Microscale, University of Science and Technology of China, Hefei, Anhui
230026, China}
\affiliation{Synergetic Innovation Center of Quantum Information and Quantum Physics,
University of Science and Technology of China, Hefei, Anhui 230026, China}

\author{NingHua~Tong} \email{nhtong@ruc.edu.cn}
\affiliation{Department of Physics, Renmin University of China, Beijing
100872, China}

\author{JianHua~Wei}
\affiliation{Department of Physics, Renmin University of China, Beijing 100872, China}

\author{YiJing~Yan} \email{yyan@ust.hk}
\affiliation{Hefei National Laboratory for Physical Sciences at the
Microscale, University of Science and Technology of China, Hefei, Anhui
230026, China}
\affiliation{Department of Chemistry, Hong Kong University of Science
and Technology, Hong Kong, China}

%\date{\today}
%\date{Submitted on September~4, 2013; resubmitted on May~21, 2014}
%\date{Submitted on September~4, 2013; revised on \today}
\date{Submitted on September~4, 2013; resubmitted on July~14, 2014}

\begin{abstract}

A nonperturbative quantum impurity solver is proposed based on a formally exact hierarchical equations of motion (HEOM) formalism for open quantum systems. It leads to quantitatively accurate evaluation of physical properties of strongly correlated electronic systems, in the framework of dynamical mean-field theory (DMFT).
The HEOM method is also numerically convenient to achieve the same level of accuracy as that using the state-of-the-art numerical renormalization group impurity solver at finite temperatures. The practicality of the novel HEOM+DMFT method is demonstrated by its applications to the Hubbard models with Bethe and hypercubic lattice structures. We investigate the metal-insulator transition phenomena, and address the effects of temperature on the properties of strongly correlated lattice systems.

\end{abstract}


%%%%%%%%%
%%%%PACS
%71.10.Fd, Hubbard Model, electronic structure
%71.27.+a, Strongly correlated electron systems
%71.30.+h, Insulator-metal transitions
%%%%%%%%%

\maketitle

\section{Introduction}

Strongly correlated electronic systems (SCS) generally refer to materials
with \emph{d-} or \emph{f-}electrons.
These localized electrons strongly interact with each other and with
the surrounding itinerant electrons, leading to various intriguing
many-body phenomena, such as Kondo phenomena, metal-insulator
transition in transition metal oxides, and heavy fermion in
rare-earth compounds.
SCS quickly becomes an active research area in condensed matter physics.\cite{Ful95}

Exact treatment of SCS for classic band theory and model methods
is difficult because of the nonperturbative many-body nature
and the complexity of SCS materials.
Since the proposal of investigating SCS in the limit of infinite
dimension in 1989,\cite{Met89324}
the dynamical mean-field theory (DMFT)\cite{Jar92168,Geo926479,Geo9613}
has been rapidly developed and applied,
leading to a dramatic progress in understanding the properties of SCS.
In this limit, the self-energy is local, so that the spatial
fluctuations in strongly correlated systems can be neglected
while only the on-site Coulomb interaction is taken into consideration.
As a result, the many-body lattice problem is mapped to an effective impurity problem,
in which the correlated electrons on the impurity site interacts
with a frequency dependent mean field.
This mean field is then updated in a self-consistent way via the
solution of the impurity model
in order to obtain the self-energy of the lattice Green's function.

The DMFT method was firstly used with Hubbard model, and revealed
the three-peak spectral function induced by the strong electron-electron correlation.\cite{Jar92168,Geo926479}
These three peaks are comprised of one central quasi-particle peak
around the Fermi energy and two single-particle Hubbard peaks.
In some transition metal oxides, the central quasi-particle peak
gradually transfers its weight to the Hubbard peaks
with increasing Coulomb interaction,\cite{Zha931666} finally
leading to metal-insulator transition,
which is usually referred to as Mott transition.\cite{Geb97}
It is one of the most fascinating phenomena of SCS that has
been observed in various transition metal oxides,\cite{Ima981039}
and extensively investigated by DMFT
methods.\cite{Geo937167,Roz9410181,Roz993498,Bul99136,Bul01045103,%
Dai05045111,Ono03035119,Flo02165111}
In recent years, the DMFT methods have been extended to the
investigations of some complicated SCS with multi orbital,
nonlocal Coulomb and exchange interaction, or taking into account
spatial correlations.\cite{Si963391,Chi003678,Kot2001,Drc0561,Kus06054713,Tos07045118}
Especially, the combination of DMFT method with density functional
theory makes it possible to simulate complicated real compounds
with strong electron correlations.\cite{Ani977359,Kot06865}
The DMFT method has already been a powerful tool for the investigation
of strongly correlated physics.

The accuracy and efficiency of DMFT calculation is determined by
the key component, the impurity solver, that solves the effective impurity problem.
A vast amount of theoretical efforts have been devoted to achieving this goal.
The perturbative methods such as the iterated perturbation
theory\cite{Kaj9616214} and non-crossing
approximation\cite{Bic87845} sum the perturbative series of
interaction diagrams to different orders, resulting in different
levels of accuracy.
Extensions of these methods
to away from half filling (iterated perturbation theory)
or to very low temperature regime (non-crossing approximation)
are technically complicated, which largely limit the applications of these methods.
The currently used nonperturbative methods include the exact diagonalization
approach,\cite{Dag94763,Caf941545,Si942761}
quantum Monte Carlo approach,\cite{Hir862521, Gul11349} and numerical
renormalization group (NRG)
method.\cite{Wil75773,Bul08395,Hof001508,Wei07076402,Pet06245114}

Although the existing methods have been successfully used in
characterizing the fundamental characteristics of SCS,
they have their own limitations.\cite{And6141,Hub63238,Lee8699}
Both the exact diagonalization approach and the
configuration interaction methods diagonalize the Hamiltonian in a
finite dimensional Hilbert space. In DMFT applications, the
continuous bath degrees of freedom are discretized and a small
number of bath sites (orbitals) are chosen to represent the full
bath. Consequently, the local density of states is composed of
discrete peaks which are then broadened artificially. The energy
resolution in the spectrum depends on the number of bath sites.
Therefore, although the quasiparticle weight $Z$ can be extracted from
the Matsubara Green's function, it is difficult to obtain the real
frequency quantities,\cite{Ani2010} including the low energy Kondo
peak and the high energy Hubbard peaks; see for instance, Fig.~S11 in
the Supplemental Material of Ref.~\onlinecite{Li12266403}.
The computational cost of some quantum Monte Carlo (QMC) approaches,
including the Hirsch-Fye algorithm and the continuous time QMC,
increases dramatically as the temperature decreases.
Moreover, numerical methods such as the maximum entropy method are
often required to extract real frequency spectral function from the
imaginary time domain via some analytic continuation operation.
This may introduce additional errors.\cite{Jar96133}
In recent years the NRG method has achieved significant progress,
thanks to the improvement of its algorithm and advancement of computer hardware.
However, some of its basic features have remained unchanged:
It utilizes logarithmic discretization and truncation of
energy spectrum during iterative diagonalization.
Therefore, quantities produced by the NRG method are less accurate
at high energy or high temperatures than at low energy or low temperatures.\cite{Bul08395}
Therefore, an accurate and efficient DMFT impurity solver
for the investigation of the strong correlation effects
at finite temperatures is highly desirable.

%%%%%%%%%%%%%%%%%%%
%{\bf
%%%%%%%%%%%%%%%%%%%

In this work we propose to use a hierarchical equations of motion (HEOM)
approach\cite{Jin08234703} as the impurity solver of DMFT.
The HEOM method treats quantum impurity systems from the perspective of open dissipative dynamics.
As will be discussed in \Sec{sec2a}, in principle the HEOM formalism is formally exact,
as long as the bath environment satisfies Gaussian statistics, which is true for noninteracting electron reservoirs.
The HEOM theory can be established based on the Feynman--Vernon path-integral formalism,\cite{Jin08234703,Fey63118,Kle09,Xu05041103,Xu07031107}
in which all the system-bath correlations are taken into consideration.

In practice, the HEOM method is implemented without invoking any approximation or any tricky controlling parameter.
The numerical results are quantitatively accurate for a wide range of impurity systems,
as long as they converge with respect to the increasing truncation level $L$.
%

%\textbf{
Usually, the HEOM results converge uniformly and rapidly with the increasing $L$.
A higher $L$ is needed to achieve numerical convergence at a lower temperature.
%}
Consequently, the computational cost increases substantially as the temperature lowers.
Currently for a symmetric single impurity Anderson model,
the HEOM approach reaches quantitatively accuracy for temperature
$T\geqslant0.1\,T_{\rm K}$ (Kondo temperature)
with the computational resources at our disposal.\cite{Li12266403}
In particular, the HEOM approach remains accurate at high temperature and large frequency (energy) range.
%
%\textbf{
For all the results presented in this work,
$L$ is chosen to be sufficiently large so that numerical convergence is always guaranteed.
%}

The HEOM approach is applicable to a general quantum impurity system.
It is capable of treating strongly correlated impurities with more than one orbitals.
The applicability of HEOM approach to various equilibrium
and nonequilibrium properties of quantum dot systems has been demonstrated.\cite{Zhe08184112,Wan13035129}
These include the studies on dynamical Coulomb blockade\cite{Zhe08093016}
and dynamical Kondo transitions in quantum dots.\cite{Zhe09164708,Zhe13086601}
%
%\textbf{
Previous studies have shown that the HEOM approach is capable of achieving the accuracy of
the latest high-level NRG approach, as demonstrated by the impurity spectral functions\cite{Li12266403}
and the local magnetic susceptibility\cite{Wan13035129} of Anderson impurity model systems at finite temperatures.
%}
All these successful applications suggest that the HEOM approach
is very suitable to be used as an impurity solver in the framework of DMFT method.

The remainder of this paper is organized as follows.
The HEOM based impurity solver is introduced in \Sec{sec2},
together with its main features and current limitations.
In \Sec{sec3}, the results for the Mott transition from DMFT using HEOM impurity solver
are systematically compared with previous NRG results
to show its accuracy and efficiency. Concluding remarks are finally given in \Sec{sec4}.

%%%%%%%%%%%%%%%
%}
%%%%%%%%%%%%%%%%

\section{Hierarchical equations of motion based DMFT impurity solver}
\label{sec2}

\subsection{A formally exact HEOM formalism for quantum impurity systems}
\label{sec2a}

The HEOM approach is based on quantum dissipation theory, which can be
used to characterize the quantum impurity systems as open systems
embedded in surrounding environments composed of itinerant electron reservoirs.
%
%{\bf
The derivation of the HEOM formalism has been detailed in Refs.~\onlinecite{Jin08234703}, \onlinecite{Zhe09164708}, and \onlinecite{Zhe121129}.
Here, we introduce some of its basic features.
%}

%%%%%%%%%%%%%%%%%%
%{\bf
%%%%%%%%%%%%%%%%%%

The Hamiltonian of the quantum impurity system can be expressed as
\begin{equation}
 H_{\T}=H_{\s}+H_{\B}+H_{\SB},
\end{equation}
where $H_{\s}$ represents the quantum impurity systems
containing strong correlation interactions,
$H_{\B} = \sum_k \ep_k \hat{d}^\dag_k \hat{d}_k$ is the noninteracting electron reservoirs, and
$H_{\SB}=\sum_{\mu k} (t_{\mu k}\, \hat{a}^\dag_{\mu}
\hat{d}_{k} + {\rm H.c.})$ accounts for the impurity-reservoir couplings.
In these terms, $\hat{a}_{\mu}^\dag$ and $\hat{a}_{\mu}$ are the
creation and annihilation operators for the impurity state $|\mu\rangle$;
$\hat{d}_{k}^\dag$ and $\hat{d}_{k}$ are those for the reservoir state
$|k\rangle$ of energy $\epsilon_{k}$.
The hybridization functions of the electron reservoir are defined as
$\Delta_{\mu\nu}(\w) \equiv \pi\sum_{k} t_{\mu k}t^\ast_{\nu k}\, \delta(\w-\ep_{k})$.
%
%%%%%%%%%%%%%%%%%%%%%%%%%%%%%%
%}
%%%%%%%%%%%%%%%%%%%%%%%%%%%%%%

In the quantum dissipation theory, the quantity of primary interest is the reduced density matrix of the impurity system, $\rho(t) \equiv {\rm tr}_{\B}\,\rho_{\T}(t)$. Here, $\rho_{\T}(t)$ is the density matrix of the total system (impurity plus electron reservoir); and ${\rm tr}_{\B}$ denotes the trace over all reservoir's degrees of freedom. The dynamics of the reduced density matrix can be characterized by a Liouville-space quantum propagator $\mathcal{U}(t,t_0)$ as follows,
\be
   \rho(t) = \mathcal{U}(t,t_0) \rho(t_0).
\ee
The specific form of $\mathcal{U}(t,t_0)$ can be expressed using the Feynmann--Vernon path integral representation of\cite{Fey63118}
\be
  \mathcal{U}(t,t_0) = \int_{t_0}^t \mathcal{D}\psi \int_{t_0}^t \mathcal{D}\psi' \,
  e^{iS[\psi]}\, \mathcal{F}[\psi,\psi']\, e^{-iS[\psi']}. \label{def-U}
\ee
Here, $S[\psi]$ is the classical action functional associated with $H_{\s}$.
For the convenience of evaluating expectation values, usually an initial factorization ansatz is adopted for the total system at $t_0 = -\infty$:
\be
   \rho_{\T}(t_0) = \rho(t_0) \otimes \rho_{\B}^{\rm eq},  \label{def-t0}
\ee
with $\rho_{\B}^{\rm eq}$ being the equilibrium reduced density matrix of reservoir environment. We emphasize that the uncorrelated total system at $t_0 = -\infty$ is chosen only as a reference for the construction of $\mathcal{U}(t,t_0)$. For the calculation of any physical quantity or process, we first solve for the fully correlated stationary state $\rho(t_0')$ at a finite time $t_0'$, and then use $\rho(t_0')$ as the initial condition for subsequent dynamics of the impurity system.

It is $\mathcal{F}[\psi,\psi']$ that accounts for the influence of the dissipative reservoir environment to the properties of the impurity, and it is thus termed as the influence functional, which has the following form:\cite{Jin08234703}
\begin{align}
  \mathcal{F}[\psi,\psi'] &\equiv \exp \left\{
  -\int_{t_0}^t d\tau \, \mathcal{R}[\tau; \psi,\psi']
  \right\}, \label{def-F} \\
  \mathcal{R}[t; \psi,\psi'] &= \sum_{\sigma\mu} a_\mu^{\bar{\sigma}}(\psi(t))
  \left[ B^\sigma_\mu(t;\psi) - B'^\sigma_\mu(t;\psi') \right] - \nl
  &\qquad \left[ B^\sigma_\mu(t;\psi) - B'^\sigma_\mu(t;\psi') \right] a_\mu^{\bar{\sigma}}(\psi'(t)), \label{def-R} \\
  B^\sigma_\mu(t;\psi) &\equiv \sum_\nu \int^t_{t_0} d\tau \,
  C^\sigma_{\mu\nu}(t,\tau)\,a^\sigma_\nu(\psi(\tau)), \label{def-B} \\
  B'^\sigma_\mu(t;\psi') &\equiv \sum_\nu \int^t_{t_0} d\tau \,
  C^{\bar{\sigma}\ast}_{\mu\nu}(t,\tau)\,a^\sigma_\nu(\psi'(\tau)),
  \label{def-Bp}
\end{align}
where $\sigma=+/-$ and $\bar{\sigma} = -\sigma$.
It should be emphasized that the above \Eq{def-R} is an exact formula, as long as the reservoir environment satisfies Gaussian statistics, which is true for reservoirs consisting of noninteracting electrons.\cite{Jin08234703}

Although $\mathcal{F}[\psi,\psi']$ has a rather complicated form, it is apparent that the impurity-reservoir coupling $H_{\SB}$ enters (exclusively) through the reservoir correlation function (or memory kernel) $C^\sigma_{\mu\nu}(t,\tau)$.
Therefore, with the exact $C^\sigma_{\mu\nu}(t,\tau)$ at hand, we can obtain the exact Liouville propagator $\mathcal{U}(t,t_0)$ and hence the exact $\rho(t)$ via the path integral of \Eq{def-U}.

For a noninteracting reservoir, the correlation function $C^\sigma_{\mu\nu}(t,\tau)$ is fully determined by the reservoir hybridization function $\Delta_{\mu\nu}(\w)$ via the following fluctuation-dissipation theorem:
\begin{align}
   C^\sigma_{\mu\nu}(t,\tau) &= \exp\left\{\bar{\sigma} i\int_\tau^t dt'\,
   V(t') \right\} \, \tilde{C}^\sigma_{\mu\nu}(t-\tau), \nl
   \tilde{C}^\sigma_{\mu\nu}(t) &= \int_{-\infty}^\infty d\w \,
   e^{\sigma i\w t} f^\sigma_\beta(\w) \Delta^\sigma_{\mu\nu}(\w). \label{def-C}
\end{align}
Here, $V(t)$ is the time-dependent bias voltage applied to the reservoir, $\Delta^+_{\nu\mu}(\w) = \Delta^-_{\mu\nu}(\w) = \Delta_{\mu\nu}(\w)$.
$f^\sigma_\beta(\w) = 1/[1 + e^{\sigma\beta(\w-\ep_f)}]$ is the Fermi function for electron ($\sigma=+$) or hole ($\sigma=-$), $\beta = 1/k_B T$, and $\ep_f$ is the equilibrium chemical potential of reservoir.
In fact, $\tilde{C}^\sigma_{\mu\nu}(t)$ are exactly equivalent to the ``embedding'' self energies: $\tilde{C}^{+}_{\mu\nu}(t) = i[\Sigma^{<}_{\mu\nu}(t)]^{\ast}$ and $\tilde{C}^{-}_{\mu\nu}(t)=i\Sigma^{>}_{\mu\nu}(t)$.\cite{Jin08234703}

Clearly, $\mathcal{U}(t,t_0)$ can be obtained via the formally exact path integral formalism by combining \Eqs{def-U}--\eqref{def-C}. The only requirement is the reservoir should consist of noninteracting electrons, and hence the reservoir environment satisfies Gaussian statistics. With the resulted $\rho(t)$, the expectation value of any system operator $\hat{A}$ can then be calculated as $\langle \hat{A} \rangle = {\rm tr} [\hat{A} \rho(t)]$.

In practice, the path integral formalism will lead to an integro-differential quantum master equation for $\rho(t)$, and hence very difficult to solve. To circumvent this problem, an HEOM formalism is instead proposed. In the HEOM formalism, the integro-differential equation is replaced by a hierarchical set of linear differential equations. As a consequence, the numerical calculations become much more convenient.

The central step towards the establishment of HEOM is the decomposition of reservoir memory kernel $\tilde{C}^\sigma_{\mu\nu}(t)$ into exponential functions, \emph{i.e.},
\be
  \tilde{C}^\sigma_{\mu\nu}(t) = \sum_{m=1}^M  \eta^\sigma_{\mu\nu m} \, e^{-\gamma^\sigma_{\mu\nu m}t}, \label{def-M}
\ee
where $1/{\rm Re}[\gamma^\sigma_{\mu\nu m}]$ is the characteristic memory time of $m$th dissipation mode.
A number of memory decomposition schemes have been developed, including the conventional Matsubara decomposition scheme,\cite{Jin08234703} a hybrid spectrum decomposition and frequency dispersion scheme,\cite{Zhe09164708} the partial fraction decomposition scheme,\cite{Cro09073102} and the Pad\'{e} spectrum decomposition scheme.\cite{Oza07035123,Hu10101106,Hu11244106}

Most of the existing memory decomposition schemes are based on a contour integral algorithm for \Eq{def-C} with the use of a residual theorem,\cite{Zhe09164708}
and each exponential term in \Eq{def-M} corresponds to a pole of complex
function $f^\sigma_\beta(z)$ or $\Delta_{\mu\nu}(z)$ in \Eq{def-C}, via the following
expansions:
\begin{align}
  f^\sigma_\beta(z) &\simeq \frac{1}{2} - \sigma \frac{1}{\beta} \sum_{p=1}^P
  \left( \frac{1}{z + z_p} - \frac{1}{z - z_p} \right), \label{pole_f} \\
  \Delta_{\mu\nu}(z) & \approx \sum_{i=1}^N \frac{\eta_{\mu\nu,i}}{(z - \Omega_{\mu\nu,i})^2 + (W_{\mu\nu,i})^2}. \label{pole_delta}
\end{align}
Therefore, the total number of memory components is $M = P + N$, with $P$ and $N$ being the
number of poles for $f^\sigma_\beta(z)$ and $\Delta^\sigma_{\mu\nu}(z)$ in the upper or lower $z$-plane, respectively.

Note that \Eq{pole_f} is a formal expansion of the Fermi function, and the
choice of $\{z_p\}$ is not unique. At any finite temperature, $P$ can always
be chosen so that the right-hand side of \Eq{pole_f} and their contributions to
$\tilde{C}^\sigma_{\mu\mu}(t)$ converge to a preset precision.
To the best of our knowledge, the Pad\'{e} expansion of $f^\sigma_\beta(z)$ requires a smallest $P$,
and thus is by far the most efficient scheme for \Eq{pole_f}.\cite{Hu11244106}
In contrast, \Eq{pole_delta} is realized by a least square fit of $\Delta_{\mu\nu}(\w)$
to a linear combination of Lorentzian functions, with $\{\eta_{\mu\nu,i}, \Omega_{\mu\nu,i}, W_{\mu\nu,i}\}$
being the fitted parameters. The accuracy of the resulted memory
decomposition is determined by the quality of the least square fit. For a continuous $\Delta_{\mu\nu}(\w)$
the fitting of \Eq{pole_delta} is usually satisfactory; see for instance \Fig{fig2} in \Sec{heom_dmft_solver}.
Therefore, the combined Pad\'{e}-Lorentzian scheme\cite{Wan13205126} achieves an accurate and efficient exponential
decomposition of $\tilde{C}^\sigma_{\mu\nu}(t)$ via \Eq{def-M}.

With the exponential decomposition of reservoir memory,
the HEOM can be cast into a compact form as\cite{Jin08234703}
\begin{align}\label{HEOM}
   \dot\rho^{(n)}_{j_1\cdots j_n} =&
   -\Big(i{\cal L} + \sum_{r=1}^n \gamma_{j_r}\Big)\rho^{(n)}_{j_1\cdots j_n}
%\nl &
     -i \sum_{j}\!     %\sideset{}{'}
     {\cal A}_{\bar j}\, \rho^{(n+1)}_{j_1\cdots j_nj}
\nl &
    -i \sum_{r=1}^{n}(-)^{n-r}\, {\cal C}_{j_r}\,
     \rho^{(n-1)}_{j_1\cdots j_{r-1}j_{r+1}\cdots j_n}.
\end{align}
Here, $\rho^{(0)}(t) = \rho(t) \equiv {\rm tr}_{\B}\,\rho_{\T}(t)$
is the reduced density matrix, and $\{\rho^{(n)}_{j_1\cdots j_n}(t); n=1,\cdots,L\}$ are the auxiliary density matrices, with $L$ being the truncation level.
The multi-component index $j\equiv (\sigma\mu \nu m)$.
The Grassmann superoperators ${\cal A}_{\bar j}\equiv {\cal A}^{\bar\sigma}_{\mu}$
and ${\cal C}_j\equiv {\cal C}^{\sigma}_{\mu\nu m}$ are
defined via their actions on a fermionic/bosonic operator $\hat{O}$
as ${\cal A}^{\bar\sigma}_{\mu} \hat O \equiv
  [\hat a^{\bar\sigma}_{\mu}, \hat O]_\mp$
and
${\cal C}^{\sigma}_{\mu\nu m} \hat O  \equiv
  \eta^{\sigma}_{\mu\nu m}\hat a^{\sigma}_{\nu}\hat O
  \pm (\eta^{\bar\sigma}_{\mu\nu m})^{\ast}\hat O\hat a^{\sigma}_{\nu}$,
respectively, with $\bar\sigma$ denoting the opposite sign of $\sigma=+/-$.
The electron correlation interaction is contained in the Liouvillian of impurities,
$\mathcal{L}\,\cdot \equiv [H_{\s}, \cdot\,]$.

Moreover, the subscript index set ($j_1\cdots j_n$) in an
$n^{\rm th}$-level auxiliary density matrix $\rho^{(n)}_{j_1\cdots j_n}$ belongs to an
ordered set of $n$ distinct $j$-indices.
The number of distinct $j$-indices, $K$, is equal to not
only the number of the first-level auxiliary density matrices in total,
but also the maximum
hierarchical level ($L_{\rm max}=K$).
Therefore, the number of the $n^{\rm th}$-level auxiliary density matrices is
$\frac{K!}{n!(K-n)!}$.
Then, the total number of unknowns to solve ${\cal N}(K,L)$, which dominates the computational cost of the HEOM approach,
is the summation over all the truncation level,
$\sum_{n=0}^{L}\frac{K!}{n!(K-n)!}\leqslant 2^K$, as $L\leqslant K$.
In this expression, $K$ can be evaluated by $K=2MN_\mu$, in which $M$ is determined by the resolution
of the bath memory, and $N_\mu$ is the number of the system states (orbitals)
that couple directly to electron reservoirs.
This scheme provides an optimal basis to
exploit the different characteristic time scales associated with
system-reservoir dissipation processes.
The hierarchy terminates automatically at $L=2$ for noninteracting $H_{\s}$;\cite{Jin08234703}
while for $H_{\s}$ involving electron correlation
interactions, the solution of \Eq{HEOM} must go through systematic
tests to confirm its convergence versus $L$.
In practice, recent investigations on a general quantum impurity system
have indicated that a relatively low $L$ ($\simeq 4$)
is usually sufficient to yield quantitatively converged results.\cite{Li12266403}

In the framework of HEOM, the density matrix of the total system $\rho_{\T}(t)$ is effectively ``folded''
into the impurity subsystem -- represented by the reduced density matrix $\rho(t) = \rho^{(0)}(t)$ along with all the
auxiliary density matrices $\{\rho^{(n)}_{j_1\cdots j_n}(t); n=1,\cdots,L\}$. The influence of
reservoir environment is fully accounted for by the reservoir memory kernel $C^\sigma_{\mu\nu}(t,\tau)$.
As long as the reservoir environment ($H_{\B}$) is noninteracting (and hence satisfies Gaussian statistics), the HEOM formalism
is formally exact, without any approximation.

In practice, the numerical results of HEOM approach achieve quantitative accuracy as long as they converge with
respect to the number of memory components $M$ (of \Eq{def-M}) and the truncation level $L$.
There is no other controlling parameter. For all the results presented in this work, the numerical convergence has been affirmed.

\subsection{Linear response theory via the HEOM dynamics}
\label{LRT}

The HEOM~\eqref{HEOM} formally define a quantum Liouville propagator $\bfG(t,\tau)$,
which associates the reduced and the auxiliary density matrices at time $t$ to those at time $\tau$:
\begin{equation}\label{pgG}
\bm{\rho}(t)=\bfG(t, \tau)\bm\rho(\tau).
\end{equation}
Here, a bold symbol denotes a quantity in the HEOM Liouville space defined by \Eq{HEOM}.
For instance, $\bm\rho(t)$ is a vector in the HEOM space and is comprised of all the
density matrices involved in \Eq{HEOM}, \emph{i.e.}, $\bm\rho(t)\equiv\big\{\rho^{(n)}_{j_1\cdots j_n}(t);\, n=0,\cdots\!,L\}$.

Denote $\bm\rho^{\rm eq}(T)\equiv
\big\{\rho^{(n);\,{\rm eq}}_{j_1\cdots j_n}(T);
\, n=0,\cdots\!,L\big\}$ as an equilibrium-state solution of \Eq{HEOM} at a given temperature $T$.
The auxiliary density matrices are nonzero, $\rho^{(n>0);\,{\rm eq}}_{j_1\cdots j_n}(T)\neq 0$,
which reflect the presence of correlations between the impurity and reservoirs.
Consider an impurity system initially at thermal equilibrium,
its dynamics starts with $\bm\rho(t_0)=\bm\rho^{\rm eq}(T)$.
In the absence of any probe field, the system equilibrium propagator satisfies the translational
invariance in time, $\bfG(t,\tau)=\bfG(t-\tau)$,
and hence $\bm{\rho}(\tau')=\bfG(\tau'-t_0)\bm\rho(t_0)=\bm\rho^{\rm eq}(T)$.

If a probe field is applied to the impurity (the perturbation Hamiltonian is $H_{\rm pr}(t)$ and the corresponding
HEOM Liouvillian is $\delta\boldsymbol{\mathcal{L}}_{\rm pr}(t)$),
the response of the reduced and auxiliary density matrices $\delta\bm\rho(t) \equiv
\big\{\delta\rho^{(n)}_{j_1\cdots j_n}(t); \, n=0,\cdots\!,L\big\}$ can be obtained formally as
\begin{widetext}
\be\label{HEOM_perturbation}
\delta{\bm \rho}(t)
 =\sum_{n=1}^{\infty}
    (-i)^n \int_{-\infty}^{t}\!\! d\tau_n \cdots\!
    \int_{-\infty}^{\tau_2}\!\!d\tau_1\,
  {\bfG}(t-\tau_n)\delta\boldsymbol{{\cal L}}_{\rm pr}(\tau_n){\bfG}(\tau_n-\tau_{n-1})
 \cdots
   \delta\boldsymbol{{\cal L}}_{\rm pr}(\tau_1)\bm\rho^{\rm eq}(T)\, .
\ee\end{widetext}

Equation~\eqref{HEOM_perturbation} is formally analogous to the celebrated Hilbert-space time-dependent perturbation
theory in the interaction picture.
Such formal analogy highlights a straightforward
equivalence mapping between the conventional Hilbert-space
and the HEOM Liouville-space formulations
for the response properties of quantum impurity systems.

Consider, for example, the equilibrium-state two-time correlation function
between two arbitrary dynamical variables $\hat A$ and $\hat B$ of the
impurity system. We have
%%%%%%%%%%
%%%%%%%%%%%%%%
\begin{align}\label{CAB}
 \check{C}_{AB}(t) =& \big\la \hat A(t)\hat B(0)\big\ra
                       =  \big\la\big\la \hat A(t) \big|
   \hat B\rho^{\text{eq}}_{\text{total}}(T)\big\ra\big\ra
\nl  =& \big\la\big\la \hat{\bm A}(t) \big|
   \hat{\bm B} {\bm\rho}^{\text{eq}}(T)\big\ra\big\ra \, .
\end{align}
Here, $\la\la \cdot | \cdot \ra\ra$ denotes inner product between vectors.
The application of \Eq{HEOM_perturbation} leads to
the evaluation of correlation and response functions
of impurity systems via the linear response theory in the HEOM Liouville space.

Regarding the linear response theory, we are primarily interested in how
system properties (such as the expectation value
of a system observable $\hat{A}$) respond to an external
perturbation in the physical subspace of the system.
Suppose the perturbation Hamiltonian assumes the
form of
\begin{equation}
   H_{\rm pr}(t)= - \hat{B} \epsilon_{\rm pr}(t), \label{LRT_HPR}
\end{equation}
where $\hat{B}$ is a Hermitian system operator,
and $\epsilon_{\rm pr}(t)$ is the time-dependent perturbative
field which is turned on from $t_0$.
The induced variation of the expectation value of a
system observable $\hat{A}$ due to the presence of $H_{\rm pr}(t)$ is
\be \delta \hat{A}(t) = {\rm tr}_{\s} \big[ \hat{A}\, \delta\rho(t)
\big] = \big\langle \big\langle \hat{\bm A}| \delta{\bm \rho}(t)
\big\rangle \big\rangle. \label{LRT_DA} \ee
Here, $\hat{\bm A} = \hat{\bm A}(0) = \{ \hat{A}, 0, \ldots, 0 \}$
has the same vector form as $\bm\rho^{\rm eq}(T)$ and $\delta\bm\rho(t)$. Note
that all $n > 0$ components of the vector $\hat{\bm A}$ are zero.

Taking the first-order term of \Eq{HEOM_perturbation}, we
arrive at
\begin{equation}
\delta\boldsymbol{\rho}(t)= -i\int_{0}^{t}d\tau\,
\boldsymbol{\mathcal{G}}(t-\tau) \,
\delta\boldsymbol{\mathcal{L}}_{\textrm{pr}}(\tau)
\boldsymbol{\rho}^{\textrm{eq}}(T).\label{LRT_1STDR}
\end{equation}
We then define the time-independent HEOM-space superoperator
$\hat{\boldsymbol{\mathcal{B}}}$ as
\begin{equation}
\hat{\boldsymbol{\mathcal{B}}} \equiv
-\delta\boldsymbol{\mathcal{L}}_{\textrm{pr}}(t)
/\epsilon_{\textrm{pr}}(t),\label{LRT_DEFB}
\end{equation}
whose action can be determined as
\begin{equation}
\hat{\boldsymbol{\mathcal{B}}}\boldsymbol{\rho}=[\hat{B},\;\boldsymbol{\rho}].\label{LRT_ACTB}
\end{equation}
Inserting Eq.\eqref{LRT_ACTB} into \eqref{LRT_1STDR}, we have
\begin{equation}
\delta\hat{A}(t)=i\int_{0}^{t}d\tau\, \big\langle \big\langle
\hat{\boldsymbol{A}}(0)|\boldsymbol{\mathcal{G}}(t-\tau)
| \hat{\boldsymbol{\mathcal{B}}} \boldsymbol{\rho}^{\textrm{eq}}(T)
\big\rangle \big\rangle \,
\epsilon_{\textrm{pr}}(\tau),\label{LRT_A2}
\end{equation}
from which we obtain the response function in the HEOM space as
\begin{equation}
\chi_{AB}(t,\tau) = i\, \big\langle \big\langle
\hat{\boldsymbol{A}}(0)|\boldsymbol{\mathcal{G}}(t-\tau)
| \hat{\boldsymbol{\mathcal{B}}}
\boldsymbol{\rho}^{\textrm{eq}}(T)\big\rangle
\big\rangle.\label{LRT_RAB}
\end{equation}
We also obtain the correlation function in the HEOM space as
\begin{equation}
\check{C}_{AB}(t,\tau) = \big\langle \big\langle
\hat{\boldsymbol{A}}(0)|
\boldsymbol{\mathcal{G}}(t-\tau) | \hat{\boldsymbol{B}}
\boldsymbol{\rho}^{\textrm{eq}}(T)\big\rangle \big\rangle
.\label{LRT_CAB}
\end{equation}
Here, the same propagator $\bfG(t-\tau)$ as that used in \Eq{pgG}
is applied to the HEOM-space vector $\hat{\boldsymbol{B}}\boldsymbol{\rho}^{\rm eq}(T) \equiv \{
\hat{B}\rho^{(n);{\rm eq}}_{j_1 \cdots j_n}; n=0,1,\ldots, L\}$.
This amounts to taking the $\hat{\boldsymbol{B}}\boldsymbol{\rho}^{\rm eq}$ as the
initial condition at time $\tau$ and then solving \Eq{HEOM} for the final state vector
at time $t$.
Then, the expectation value of the system observable $\hat{A}$ is evaluated with such a final state.
By setting $\tau=0$ while noting
$\hat{\boldsymbol{A}}(t)=\hat{\boldsymbol{A}}(0)\boldsymbol{\mathcal{G}}(t)$,
\Eq{LRT_CAB} recovers \Eq{CAB} immediately.

In this work, the dynamical observables of our primary interest
are Green's function and spectral function of the impurity.
The spectral function is associated with the system correlation function
through the fluctuation-dissipation theorem at thermal equilibrium.
Consider the retarded single-electron Green's functions in terms of the correlation functions,
\begin{align}
 G_{AB}^{r}(t) =& -i\theta(t) \big\langle \big\{\hat{A}(t),\:\hat{B}\big\} \big\rangle
\nl =& -i\theta(t) \left[\check{C}_{AB}(t)+\check{C}_{BA}(-t)\right],
\label{RGF_def}
\end{align}
where the Green's function is defined for two arbitrary
operators $\hat{A}$ and $\hat{B}$.
We now focus on cases in which $\hat{A}=\hat{B}^\dag$ (\emph{e.g.},
$\hat{A}=\hat{a}$ and $\hat{B}=\hat{a}^\dag$ for conventional fermionic Green's
functions). In such circumstances,
$\check{C}_{AB}(t)=[\check{C}_{B^\dag A^\dag}(-t)]^\ast$ follows by definition.
We can further define the spectral function $J_{AB}(\omega)$ as
\be J_{AB}(\omega)\equiv -\frac{1}{\pi}\, \mathbf{\mathrm{Im}}
\left[G_{AB}^{r}(\omega)\right],\label{SecF_def}
\ee
in which $G_{AB}^{r}(\omega)$ is obtained by Fourier transform of $G_{AB}^{r}(t)$%

To summarize, within the framework of HEOM the system Green's functions are obtained through the following procedures:
(1) Solve HEOM~\eqref{HEOM} for the equilibrium state vector $\bm\rho^{\rm eq}(T)$ consisting of all the reduced and auxiliary density matrices.
(2) Construct the HEOM-space vector $\hat{\bm B} \bm\rho^{\rm eq}$.
(3) Propagate the HEOM~\eqref{HEOM} from $\tau=0$ to time $t$ by using the vector $\hat{\bm B} \bm\rho^{\rm eq}$ as the initial condition.
(4) Evaluate the expectation value of the system observable $\hat{A}$ to obtain the correlation function $\check{C}_{AB}(t-\tau)$.
(5) The system Green's functions are obtained by combining the correlation functions, such as via \Eq{RGF_def}.
Clearly, the evaluation of system Green's functions is exactly based on the linear response theory, and
no approximation is made throughout the above procedures.

\subsection{HEOM evaluation of dynamical quantities}

The system dynamical observable for strongly correlated electronic systems,
such as the spectral function $A(\omega)\equiv J_{\hat{a}\hat{a}^\dag}(\omega)$,
can also be evaluated separately for each frequency point.
This is achieved by the half-Fourier transform of the correlation function via,
\begin{align}
  \bar{C}_{AB}(\w) &\equiv \int_0^{\infty} dt\, \check{C}_{AB}(t)\, e^{i\w t}
  \nl &= \int_0^{\infty} dt\, \langle\langle \hat{\bm A} |
  \, \boldsymbol{\mathcal{G}}(t) | \hat{\bm B}
  \bm \rho^{\rm eq}(T) \, \rangle \rangle \, e^{i\w t} \nl
  &= \int_0^{\infty} dt \, \langle\langle \hat{\bm A} |
  e^{(i\w - \bm \Lambda) t} | \hat{\bm B} \bm \rho^{\rm eq} \rangle\rangle \nl
  &= - \langle \langle \hat{\bm A} | \big(i\w - \bm \Lambda \big)^{-1}
  | \hat{\bm B} \bm\rho^{\rm eq}  \rangle \rangle
  \equiv \langle \langle \bm A | \bm X \rangle \rangle. \label{CABw}
\end{align}
Here, the HEOM propagator is formally cast into the form of
$\boldsymbol{\mathcal{G}}(t) = e^{-\bm\Lambda t}$.
At a fixed frequency $\w$, the HEOM-space vector $\bm X$
is determined by solving the following linear problem
with the quasi-minimal residual algorithm:\cite{Fre91315,Fre93470}
\be
   \big(i\w - \bm \Lambda \big) \bm X = - \hat{\bm B} \bm\rho^{\rm eq}.
   \label{linX}
\ee
The system spectral function is then
\be\label{A_mu_freq}
  A_\mu(\w) = \frac{1}{\pi} \, {\rm Re} \Big[
  \bar{C}_{\hat{a}_\mu \hat{a}_\mu^\dag}(\w) +
  \bar{C}_{\hat{a}_\mu^\dag \hat{a}_\mu}(-\w) \Big].
\ee
Usually, the band width of the total system is finite.\cite{Ani2010}
Consequently, a finite number of frequency points within the bandwidth are sufficient
to accurately characterize the system observables.
In this work, the number of frequency points used ranges from 64 to 128.
Moreover, calculations on each individual frequency points are parallelized
to improve the efficiency of HEOM solver.

The retarded Green's function of the impurity system can be evaluated
as follows.
\be
  G^{\rm imp}_{\mu}(\w) =
  -i \left\{ \bar{C}_{\hat{a}_\mu \hat{a}_\mu^\dag}(\w)
  + [\bar{C}_{\hat{a}_\mu^\dag \hat{a}_\mu}(-\w)]^\ast \right\}.
  \label{Grw}
\ee
Apparently, one can recast \Eq{A_mu_freq} as
\begin{equation}
A_{\mu}(\w)=-\frac{1}{\pi}{\rm Im}[G^{\rm imp}_{\mu}(\w)].
\end{equation}

\subsection{HEOM for a DMFT impurity solver} \label{heom_dmft_solver}

The simplest model to investigate the correlation effect in lattice systems
is the one-band Hubbard model
\begin{equation}
H_{\rm lattice}=-\sum_{ij\sigma}t^{}_{ij}(c^\dagger_{i\sigma}c^{}_{j\sigma}
  +c^\dagger_{j\sigma}c^{}_{i\sigma})
  + U\sum_i c^\dagger_{i\uparrow}c^{}_{i\uparrow}c^\dagger_{i\downarrow}c^{}_{i\downarrow},
\end{equation}
where $c^\dagger_{i\sigma} (c_{i\sigma})$ represents creation (annihilation)
of an electron on site $i$ with spin $\sigma$,
and $t_{ij}$ is the hopping parameter between site $i$ and $j$.
Under the approximation of infinite dimension, the hopping parameter $t$
is scaled as $t=\tilde{t}/\sqrt{Z}$
with $Z$ being the lattice dimension or the number of nearest neighbors
and $\tilde{t}$ a constant.
Meanwhile, the electron-electron self-energy becomes local,
which is represented by $\Sigma^{\rm loc}_{\rm ee}(\omega)$.
Then the local Green's function of the lattice model is
\be
 G^{\rm loc}(\w)=\sum_k \frac{1}{\left[G^{0}_k(\w)\right]^{-1}
  - \Sigma^{\rm loc}_{\rm ee}(\w)},
%&=\sum_k \frac{1}{\w + i\eta - \epsilon_k -  \Sigma^{\rm loc}_{\rm ee}(\w)}.
\ee
where $G^{0}_k(\w)=(\w+i\eta-\epsilon_k)^{-1}$,
with $\eta$ being a positive minimal,
is the noninteracting lattice Green's
function component of the specified energy $\epsilon_k$.
Considering the corresponding density of states
$A_0(\epsilon)=\sum_k \delta(\epsilon-\epsilon_k)$,
the above local Green's function can be also calculated
by the integration over energy variable
\begin{equation}
G^{\rm loc}(\w)=\int d\epsilon\frac{A_0(\epsilon)}
 {\w +i\eta - \Sigma^{\rm loc}_{\rm ee}(\w) -\epsilon}.
\end{equation}

The locality of the lattice Green's function makes it possible to map the lattice problem
to effective impurity problem with exactly the same on-site electron-electron self-energy.
\cite{Kot2001}
The Hamiltonian of the single impurity model is expressed as
\begin{align}\label{H_imp}
 H^{\rm imp} &=\sum_{\sigma}\epsilon\, a^\dagger_\sigma a^{}_\sigma
  + U a^\dagger_{\uparrow}a^{}_{\uparrow}a^\dagger_{\downarrow}a^{}_{\downarrow} \nl
&\quad +\sum_{k\sigma}\left[\epsilon_k d^\dagger_{k\sigma}d_{k\sigma}
   + t_k\left(d^\dagger_{k\sigma}a^{}_\sigma +a^\dagger_\sigma d^{}_{k\sigma} \right)
    \right].
\end{align}
In the HEOM formalism the effects of coupling reservoir dynamics,
as from the last term of \Eq{H_imp},
are accounted for by the hybridization function $\Delta(\w)$.
This is just the imaginary part of the reservoir
self-energy, $\Sigma^{\rm imp}_{\rm res}(\w)$,
that can also be evaluated from the noninteracting Green's function
$G^{\rm imp}_0(\w)=[\w + i\eta -\Sigma^{\rm imp}_{\rm res}(\w)]^{-1}$.
The impurity Green's function is given by
$[G^{\rm imp}(\w)]^{-1}=[G^{\rm imp}_0(\w)]^{-1} - \Sigma^{\rm imp}_{\rm ee}(\w)$,
where $\Sigma^{\rm imp}_{\rm ee}(\w)$ denotes electron-electron self-energy.
By mapping the lattice model to an effective impurity model,
the following relations should be satisfied:\cite{Ani2010}
\begin{equation}
G^{\rm loc}(\w)=G^{\rm imp}(\w),\quad \Sigma^{\rm loc}_{\rm ee}(\w)
 =\Sigma^{\rm imp}_{\rm ee}(\w).
\end{equation}
In this way the self-consistent equations can be derived as
\be\begin{split}
 & G^{\rm loc}(\w)=\int d\epsilon\frac{A_{0}(\epsilon)}
   {\w + i\eta - \Sigma_{\rm ee}(\w) -\epsilon} = G^{\rm imp}(\w), \\
 & {[G^{\rm imp}_0(\w)]}^{-1} = [G^{\rm imp}(\w)]^{-1}+ \Sigma^{\rm imp}_{\rm ee}(\w).
\end{split}\ee

\begin{figure}
\includegraphics[width=1.0\columnwidth]{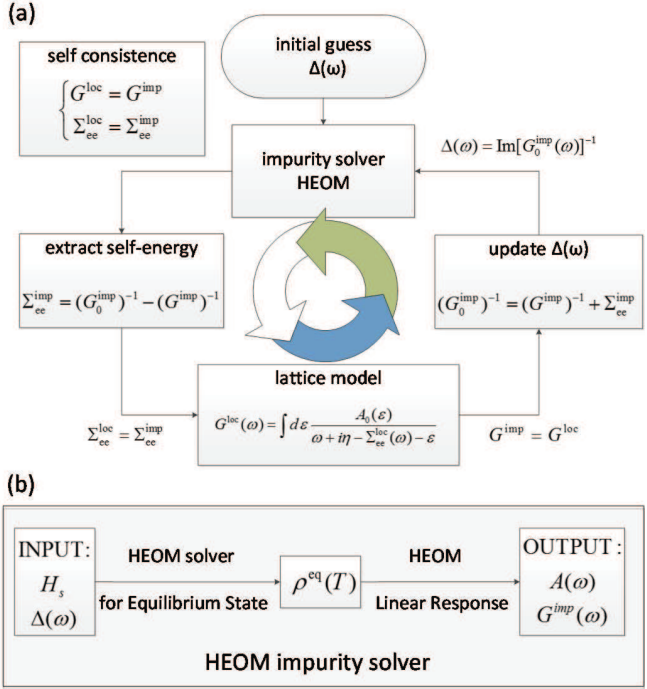}
\caption{(Color online). (a) Flow chart of the HEOM based DMFT approach. (b) Schematic diagram of the HEOM impurity solver.}
\label{fig1}
\end{figure}

The overall framework of HEOM-based DMFT is presented in the flow chart of \Fig{fig1}(a),
while the numerical procedures involved in the HEOM impurity solver is sketched in \Fig{fig1}(b).
The HEOM impurity solver starts with the pre-determined system Hamiltonian $H_s$
and the input hybridization function $\Delta(\omega)$.
First, the HEOM of \Eq{HEOM} is solved
to obtain the equilibrium-state reduced density matrix
and auxiliary density matrices $\bm\rho^{\rm eq}(T)\equiv
\big\{\rho^{(n);\,{\rm eq}}_{j_1\cdots j_n}(T);
\, n=0,\cdots\!,L\big\}$.
Then, the system correlation functions can be obtained by solving
the HEOM-space linear problem of \Eq{linX} for each frequency point.
Finally, the impurity spectral function and Green's function are evaluated
through \Eqs{A_mu_freq} and \eqref{Grw}, which are used to
construct a new hybridization function via the lattice Green's function.
%%%%%%

For achieving the optimal efficiency of HEOM,
a multi-Lorentzian decomposition scheme\cite{Zhe10114101,Xie12044113}
for the hybridization function $\Delta(\omega)$ is adopted.
%%%%%%%%%%%
%\textbf{
%%%%%%%%%%%
In this way, the hybridization function is spanned by a set of
Lorentzian functions
$\Delta(\omega)\equiv\sum_{i=1}^{N}\Delta_i(\omega)=\sum_{i=1}^{N}\frac{\Gamma_iW_i^2}{(\omega-\Omega_i)^2+W_i^2}$.
The parameters $\Omega_i, \Gamma_i$ and $W_i$ for each Lorentzian function are obtained by a least square fit.
%%%%%%%%%%
%}
%%%%%%%%%%
As the hybridization function is updated in each DMFT iteration,
the number of the Lorentzian functions $N$
is tuned case by case to find the minimal $N$ with sufficient fitting accuracy.

\begin{figure}
\includegraphics[width=1.0\columnwidth]{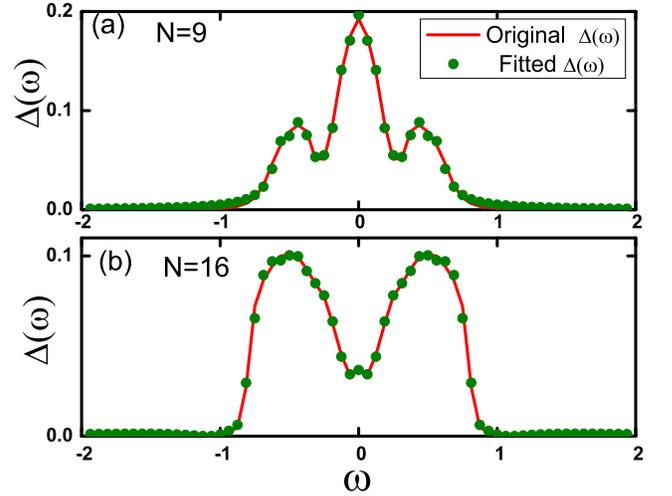}
\caption{(Color online). The Lorentzian fit of the hybridization function.
The number of Lorentzian functions used is $N=9$ for (a) and $N=16$ for (b).}
\label{fig2}
\end{figure}

As shown in \Fig{fig2}, two typical hybridization functions
are fitted by Lorentzian functions.
For a hybridization function with a well-defined peak structure,
$N=9$ is sufficient to reach a reasonable accuracy; see \Fig{fig2}(a).
However, for a hybridization function possessing a complex peak structure,
a larger $N$ is necessary to reach the same level of fitting accuracy.
As shown in \Fig{fig2}(b), the presence of a small and narrow peak
at the Fermi energy requires $N = 16$ for an accurate fitting.
At present, the major bottleneck of the HEOM approach
is the physical memory required to store all the
auxiliary density matrices (rather than the CPU time),
with the total number of ${\cal N}(K,L)=\sum_{n=0}^{L}\frac{K!}{n!(K-n)!}$,
where $K=4 M$, as described earlier (see Sec.\ \ref{sec2a}) for single impurity systems,
with $M$ being the number of the Lorentzian function
plus the number of Pad\'{e} decomposition term for the Fermi function.
For a typical calculation in this work,
the maximum memory required using Lorentzian number $N = 9$ is about 2361\,MB,
while the memory required for $N = 18$ is 11366\,MB (see Table \ref{tab1}).

\begin{figure}
\includegraphics[width=1.0\columnwidth]{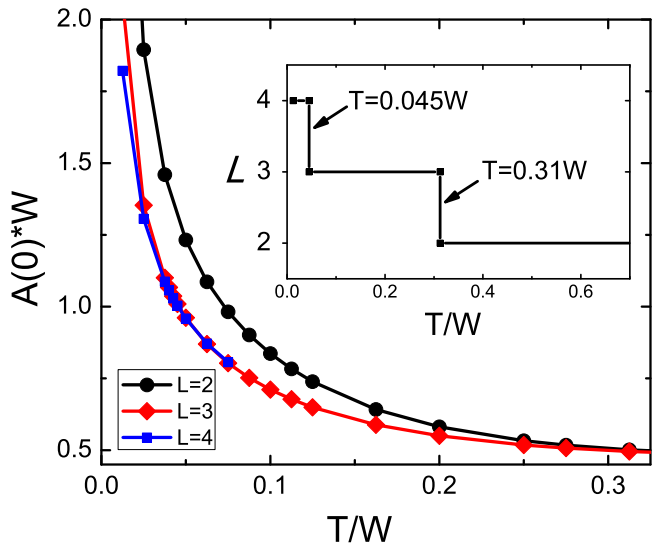}
\caption{(Color online). Spectral function $A(\omega=0)$ as a function of temperature for truncation level $L=2\sim4.$
   The inset shows the minimal $L$ that is necessary to achieve quantitative convergence at different temperatures.}
\label{fig3}
\end{figure}
The nonperturbative feature of HEOM ensures it is applicable to
SCS of a wide range of system parameters.
It goes with a well-defined convergence scheme determined by the truncation level $L$,
without any tricky parameters to deal with.
The convergence of the dynamical quantities for quantum impurity systems
is guaranteed once convergence with respect to
truncation level $L$ is reached.
%%%%%%%
%\textbf{
The minimal truncation level $L$ required to achieve convergence is closely dependent on
the system configuration and the surrounding environment, such as the strength of electron correlation,
the system-reservoir coupling, and the temperature.
It is difficult to have an a priori estimation for the required minimal $L$.
In practice, the convergence of HEOM with respect to $L$ is tested case by case.
In \Fig{fig3}, we demonstrate such a convergence test for different temperatures.
The spectral function $A(\omega=0)$ is monitored for various combinations of $L$ and $T$.
For the lowest temperature $T=0.0125W$ involved in this work, $L=4$ is necessary
(data for $L=5$ are not shown).
As the temperature increases to $T=0.045W$ and $T=0.31W$, $L=3$ and $L=2$ are sufficient
for convergence, respectively.
In this work, all the results presented are affirmed as already converged at $L \leqslant 4$.
%%%%%%%%%%
%}
%%%%%%%%%

%
In fact, the HEOM solver retains its accuracy and becomes even more efficient
at high temperatures.
In practice, the HEOM solver is applicable to an arbitrary finite temperature,
provided with sufficient computational resources.
Therefore, it is possible to investigate temperature sensitive
electron correlation effects by continuously varying the temperature.
Furthermore, the HEOM approach works in the real frequency domain,
and hence avoids invoking the numerically ill-defined analytical continuation
problems for approaches working in the imaginary frequency domain.

\begin{table}
\centering
\begin{tabular}{|r|r|r|r|}
  \hline
  $\qquad L$ & $\qquad N$ & CPU time (s)& Memory (MB)\\\hline
  1 & 9 & 2 & 0.2 \\
  2 & 9 & 34 & 9 \\
  3 & 9 & 959 & 175 \\
  4 & 9 & 48934 & 2361 \\
  5 & 9 & 302174 & 21378 \\
  4 & 14 & 99356 & 5584 \\
  4 & 18 & 1637256 & 11366 \\
  \hline
\end{tabular}
\caption{CPU time (in unit of second) and memory (in unit of megabyte)
used for typical HEOM calculations on a PC with a 3.0GHz frequency CPU.
$L$ is the truncation level, and $N$ is number of the Lorentzian functions used
for the decomposition of system-reservoir hybridization function.}
\label{tab1}
\end{table}

The major limitation of the current HEOM implementation is that the computational
cost increases dramatically as the system temperature decreases.
This is because for a lower temperature, a larger number of memory components ($M$ in \Eq{def-M})
and a higher truncation level ($L$) are necessary to ensure numerical convergence,
leading to a rapid growth of the total number ${\cal N}(K,L)$
of auxiliary density matrices involved in the HEOM formalism.
In Table \ref{tab1}, we collect the CPU time and memory usage information for some
typical HEOM impurity solver calculations.
In practical calculations, parallel programming techniques have been employed to reduce
the computational time.
The lowest temperature we are able to access with the computational resources at
our disposal is $T=0.0125W$,
in which $W$ is the effective bandwidth of the noninteracting lattice density of states.
Actually, the current HEOM implementation is mainly
limited by the insufficient physical memory to store all the auxiliary density matrices
(rather than CPU time), as mentioned earlier.
Fortunately, it is possible to improve substantially the efficiency of
HEOM by removing part of the auxiliary density matrices those are exactly zero
based on physical consideration.
This forthcoming improvement may realize the investigations of SCS
using the HEOM based DMFT at further lower temperatures.

\section{results and discussions}\label{sec3}

The Mott metal-insulator transition is one of the best test beds
for the accuracy and efficiency of the DMFT implementation.
In this work we focus on the Mott transition
in the half-filling paramagnetic phase at finite temperature.
%
%\textbf{
Employing our developed DMFT approach using the HEOM impurity solver,
we calculate the spectral functions of SCS, and compare our results with those obtained by Bulla {\it et al.}\cite{Bul99136,Bul01045103}
who used an NRG impurity solver along with a ``patching method'' for the evaluation of spectral functions.
%
%}

The calculations are performed on both the hypercubic lattice with infinite dimension
and the Bethe lattice with infinite coordination number.
The noninteracting density of states for these two lattices are
\begin{equation}
A_0^{\rm h}(\epsilon)=\frac{1}{\tilde{t}\sqrt{2\pi}}
 \exp(-\frac{\epsilon^2}{2{\tilde{t}}^2}),
\end{equation}
and
\begin{equation}
A_0^{\rm B}(\epsilon)=\frac{1}{2\pi{\tilde{t}}^2}\sqrt{4{\tilde{t}}^2-\epsilon^2};
\qquad  |\epsilon|\leqslant2\tilde{t}.
\end{equation}
The effective bandwidth, defined as
$W \equiv 4[{\int d\epsilon A_0(\epsilon)\epsilon^2}]^\frac{1}{2}$,
is calculated to be $W=4\tilde{t}$ for both
the hypercubic and the Bethe lattices.\cite{Bul99136}
In this work, $\tilde{t}$ is set to unity, and $W=4$ is set as the energy scale.
To ensure the half-filling of the impurity system, a symmetric Anderson
impurity model is employed,
and the chemical potential is set to zero.

\begin{figure}
\includegraphics[width=1.0\columnwidth]{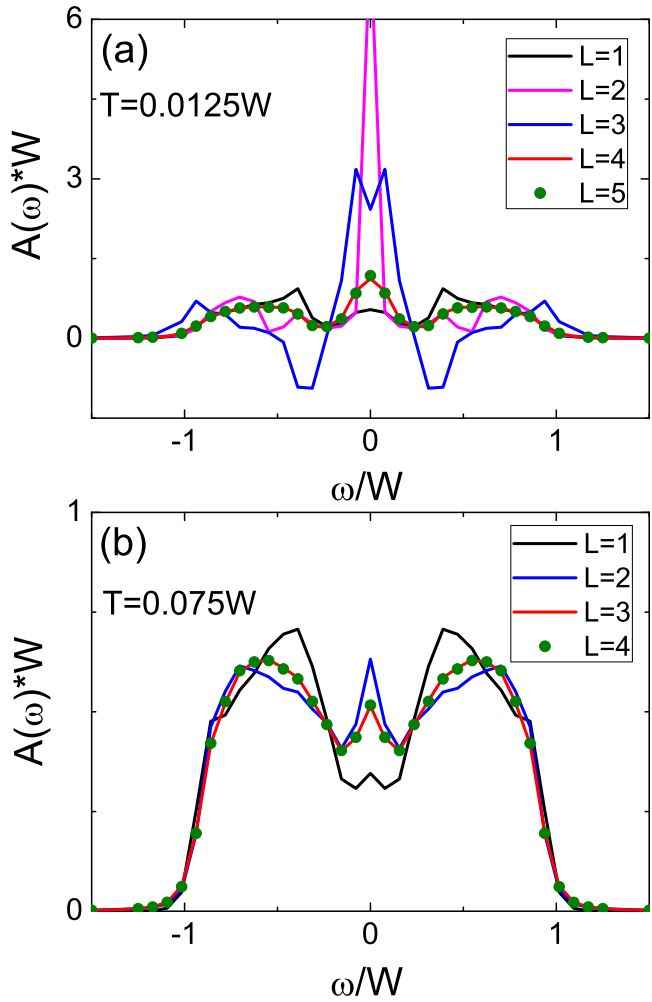}
\caption{(Color online). The convergence of the spectral function
with respect to the truncation level $L$
for Bethe lattice at (a) $T=0.0125W$ and (b) $T=0.075W$.
Here, $U=W$ with $W$ being the effective bandwidth.}
\label{fig4}
\end{figure}

The convergence of the DMFT method using the HEOM impurity
solver versus the truncation level $L$ is examined, and shown in \Fig{fig4}.
At the lowest temperature performed in our study, $T=0.0125W$,
the calculated spectral function
is converged at $L=4$.
At a truncation level lower than ($L<4$), the resulted $A(\w)$
are not accurate enough, even qualitatively.
Sometimes even unphysical results may appear,
such as the negative spectral function at $L=3$.
The typical time for a single DMFT iteration at truncation level $L=4$
is about 6 hours on a PC with a 3.0GHz frequency CPU.
When the temperature increases to $T=0.075W$, $L=3$ is sufficient for the convergence
of the resulting spectral function.
At this truncation level, the typical time for a single DMFT iteration dramatically decreases to less than 30 minutes.
In this work, the convergence of all the data represented has been carefully checked.

\begin{figure}
\includegraphics[width=1.0\columnwidth]{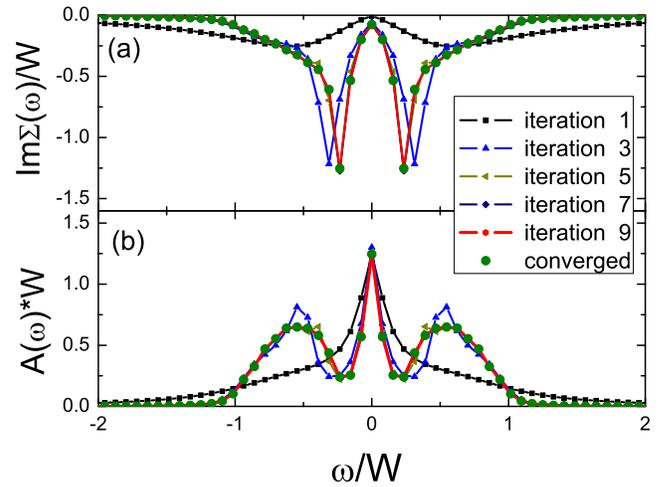}
\caption{(Color online). The convergence of (a) the self-energy and
(b) the spectral function for a hypercubic lattice by iteration.
The parameters are $U=W$ and $T=0.0125W$, where $W$ is the effective bandwidth.}
\label{fig5}
\end{figure}

The DMFT iteration, calculated by using the proposed method,
is usually converged within 10 iterations for systems
away from the metal-insulator transition region.
As an example, we show in \Fig{fig5} the convergence of
the spectral function and self-energy in hypercubic lattice
by iteration, with the parameter $U=W$ and $T=0.0125W$.
The initial hybridization function can be obtained by
setting the initial electron-electron self-energy to zero.
Alternatively, in this work we choose to use the initial hybridization
in the form of a single Lorentzian function centered at the Fermi level.
We have confirmed that they converge to the same final self-energy and spectral function.
By adopting the single Lorentzian type initial hybridization function,
numerical convergence is reached within 10 DMFT iterations.
However, more iterations are necessary to achieve convergence
when the system is near to the phase transition area.

\begin{figure}
\includegraphics[width=1.0\columnwidth]{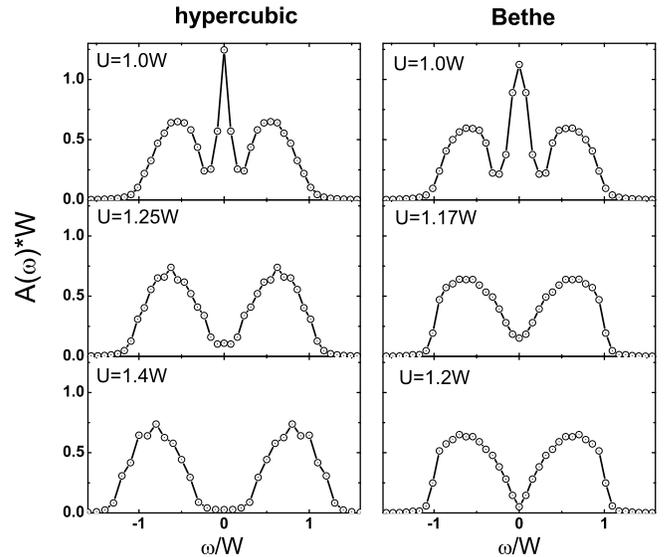}
\caption{Spectral functions of the hypercubic lattice and the Bethe lattice with
different values of $U$. The temperature is $T=0.0125W$,
where $W$ is the effective bandwidth.}
\label{fig6}
\end{figure}

We report in \Fig{fig6} the spectral functions $A(\omega)$,
with different values of $U$ at finite temperature $T=0.0125W$,
for both the hypercubic lattice and the Bethe lattice.
These two lattices have similar behaviors.
For $U=W$ both show
metallic three-peak characteristics, with a quasi-particle peak at the Fermi level.
Further increasing the values of $U$, the height of quasi-particle peak
decreases abruptly to a nearly zero value at a critical $U_c$ (see the $T = 0.0125W$ line
in the inset of \Fig{fig9}),
leading to a transition from metal to insulator.
The transition behavior is very similar to
the zero temperature spectral function,\cite{Bul99136}
except for that
the quasi-particle peaks are slightly broadened and lowered at finite temperature.

\begin{figure}
\includegraphics[width=1.0\columnwidth]{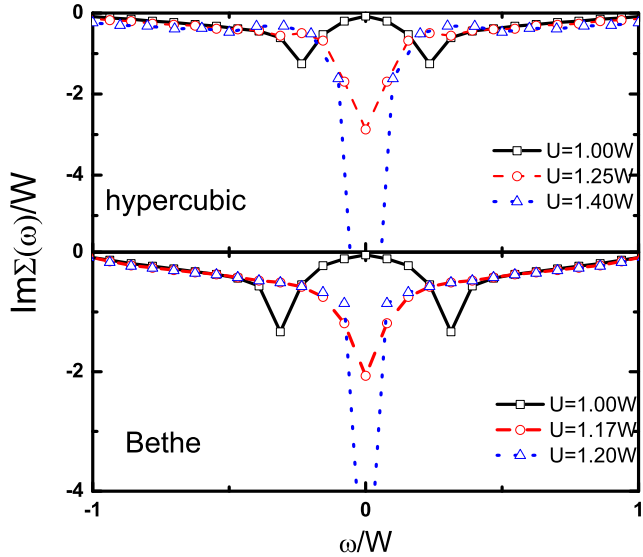}
\caption{(Color online). The imaginary part of the self-energy of
hypercubic (upper panel) and Bethe (lower panel) lattices at
different values of $U$. The temperature is $T=0.0125W$,
where $W$ is the effective bandwidth. }
\label{fig7}
\end{figure}

The corresponding converged self-energies are shown in \Fig{fig7}
for both the hypercubic and the Bethe lattice systems at the same temperature.
For $U=W$, the two-peak structure of the imaginary part
of the self-energy gives rise to the three-peak structure in the spectral function.
This is because the peaks in the imaginary part of the self-energy
induce dips in the spectral function.
With the increase of the value of $U$, the peaks in the imaginary self-energy
around the Fermi energy
show up, leading to the vanishing of the quasi-particle peaks in the spectral function.
The peaks around the Fermi energy in the imaginary self-energy are
also broadened by the finite temperature.
Thus the critical value of $U$ for metal-insulator transition would
not depend on temperature.\cite{Bul01045103}

\begin{figure}
\includegraphics[width=1.0\columnwidth]{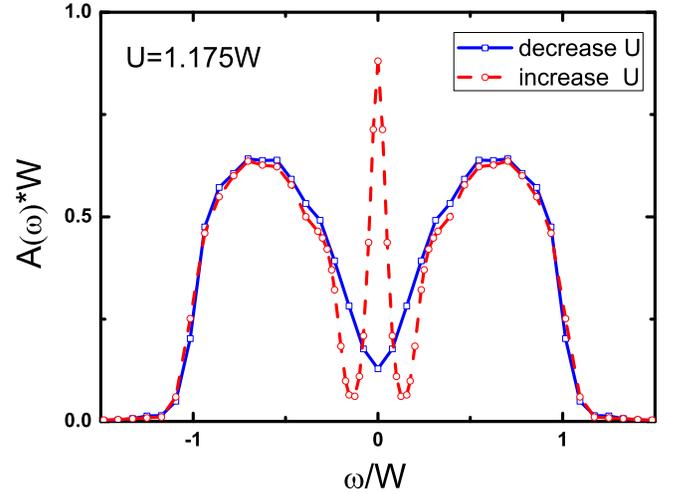}
\caption{(Color online). Existence of dual solution for the spectral function
at $T=0.0125W$ and $U=1.175W$. The two solutions are obtained
by increasing and decreasing the value of $U$ gradually to $1.175W$, respectively.
} \label{fig8}
\end{figure}

For the numerical aspect, the computational cost (time and memory) of present HEOM approach
grows rapidly with the lowered $T$ as higher truncation level is needed for convergence.
Currently the lowest temperature that is available for the present HEOM
algorithm as DMFT impurity solver
is $T=0.0125W$.

As will be shown in the following part of this section,
this currently HEOM achievable temperature is already lower
than the critical temperature $T_{\rm c}\approx0.02W$
for first-order Mott-Hubbard metal-insulator transition of
the Bethe lattice.\cite{Bul01045103}
One of the main feature for $T<T_{\rm c}$ is the coexistence of both
the metallic and insulating solutions
within a certain range of $U$.\cite{Bul01045103,Fel04136405}
This phenomenon is also found in our calculation, as shown in \Fig{fig8}.
When $U=1.175W$ at $T=0.0125W$, a metallic three-peak spectral function is obtained
if the initial spectral function is also metallic.
In this case, the converged spectral function of a lattice system with a smaller $U$
is used as the initial guess.
Whereas if a converged insulating spectral function corresponding to a lattice
system with a larger $U$
is feeded as the initial guess,
the resulted spectral function has a dramatically depressed component
around the Fermi energy.
It is noted that for the ``insulating'' solution in \Fig{fig8},
the spectral function around the Fermi energy is nonzero although the value is very small.
This is possibly because that the finite temperature of $T=0.0125W$ is very
close to $T_{\rm c}$.

One of the advantages of the HEOM approach is
that it is even more convenient to achieve quantitative accuracy
at higher temperatures.
This feature ensures the continuous variation of the temperature
to investigate possible temperature sensitive phase transitions.
It should be noted that when the temperature increases,
the efficiency of the HEOM approach is dramatically enhanced.
For example, the calculation speed for single DMFT iteration at $T=0.075W$
is more than ten times faster
than that at $T=0.0125W$.
Figure~\ref{fig9} depicts the spectral functions $A(\omega)$ and
both the real and imaginary parts of the self-energies of Bethe lattice.
The value of $U$ is fixed to $U=W$, while the temperature is varied
from $T=0.0125W$ to $T=0.1W$.
At $T=0.0125W$, the spectral function has the typical three-peak feature.
When the temperature increases, the quasi-particle peak shrinks continuously.
This change is also reflected in the corresponding self-energy.
For example, by increasing the temperature,
the two peaks in the imaginary part of the self-energy at low temperature
become smaller and move toward the Fermi energy.
Finally, these peaks disappear at high temperature $T=0.1W$, and a peak around
the Fermi energy shows up.

\begin{figure}
\includegraphics[width=0.9\columnwidth]{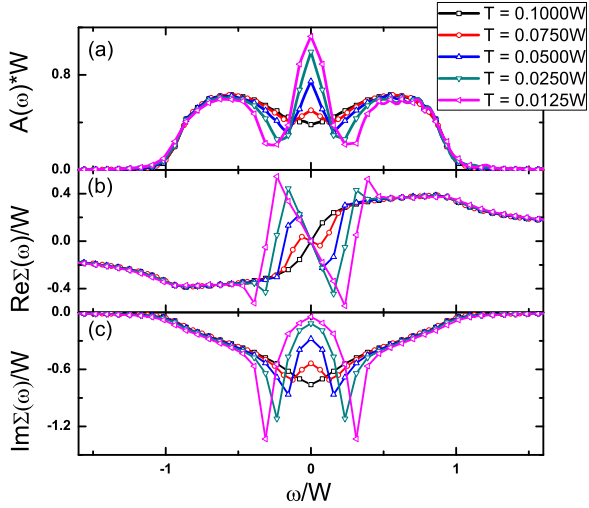}
\caption{(Color online). (a) Spectral function $A(\w)$,
(b) real part of self-energy, and (c) imaginary part of self-energy
at different temperatures for a Bethe lattice.
Here, $U=W$ with $W$ being the effective bandwidth.
}
\label{fig9}
\end{figure}

Although the quasi-particle peak shrinks in the spectral function as shown in \Fig{fig9},
the system of $U=W$ remains in the metallic state.
However, by tuning the values of $U$, metal-insulator transition may occur as the
temperature varies.

\begin{figure}
\includegraphics[width=\columnwidth]{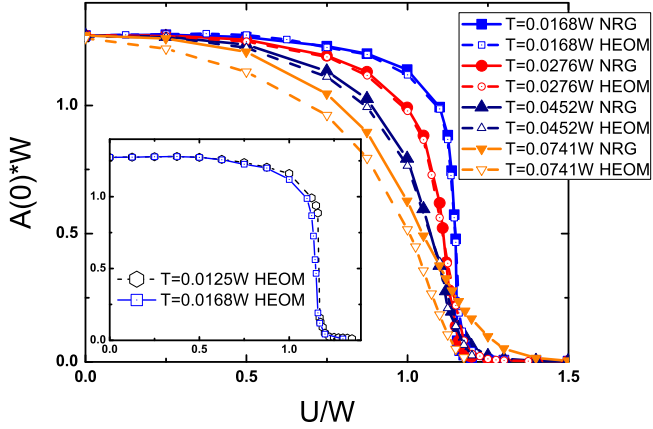}
\caption{(Color online). Comparison of $A(\w=0)$ vs $U$ between DMFT
methods using HEOM and NRG impurity solvers. Various temperatures are examined.
The inset shows the HEOM obtained $A(\w=0)$ versus $U$ for temperature $T=0.0168W$
and $T=0.0125W$.
The NRG data are extracted from Ref.~\onlinecite{Bul01045103}. } \label{fig10}
\end{figure}

In the following, we focus on the $U$ dependence of the spectral
function at the Fermi energy $A(\w = 0)$
at different temperatures, and compare to the results obtained by the NRG
method.\cite{Bul01045103}
The comparison is shown in \Fig{fig10}. At some low temperatures, for instance, $T=0.0168W$ or $T = 0.0276W$,
the HEOM results agree quantitatively with those obtained by the NRG method.
As $U$ increases from zero, $A(\w = 0)$ decreases to zero at a certain value of $U$. Such a value depends on the temperature.
Moreover, as shown in the inset, while the HEOM calculated
$A(\w = 0)$ gradually decreases to zero at $T=0.0168W$,
at $T=0.0125W$ the curve exhibits an abrupt decrease around $U=1.17W$.
This indicates a first-order phase transition at $T=0.0125W$.

As the temperature increases further, the HEOM results start
to deviate from the NRG data; see \Fig{fig10}.
The deviation is one-sided, meaning that the HEOM calculated $A(\w = 0)$
is always smaller than that obtained by the NRG method,
and the magnitude of deviation increases consistently versus temperature.

It has been elaborated in the Sec.~\ref{sec2} that the HEOM approach is in principle numerically exact.
Meanwhile, the NRG method is also considered as a numerically exact impurity solver.
Therefore, it is intriguing why the two in-principle exact approaches would give different results at the somewhat high temperature $T=0.0741W$.
To clarify this issue, we emphasize here that the numerical exactness of a certain approach
is achieved only when the computation results converge quantitatively with respect to all the involving parameters.

To understand the discrepancy between the HEOM and NRG results obtained at the high temperature,
we now check the numerical convergence of both impurity solvers,
for the same system and with the same hybridization function.

\begin{figure}
\includegraphics[width=0.95\columnwidth]{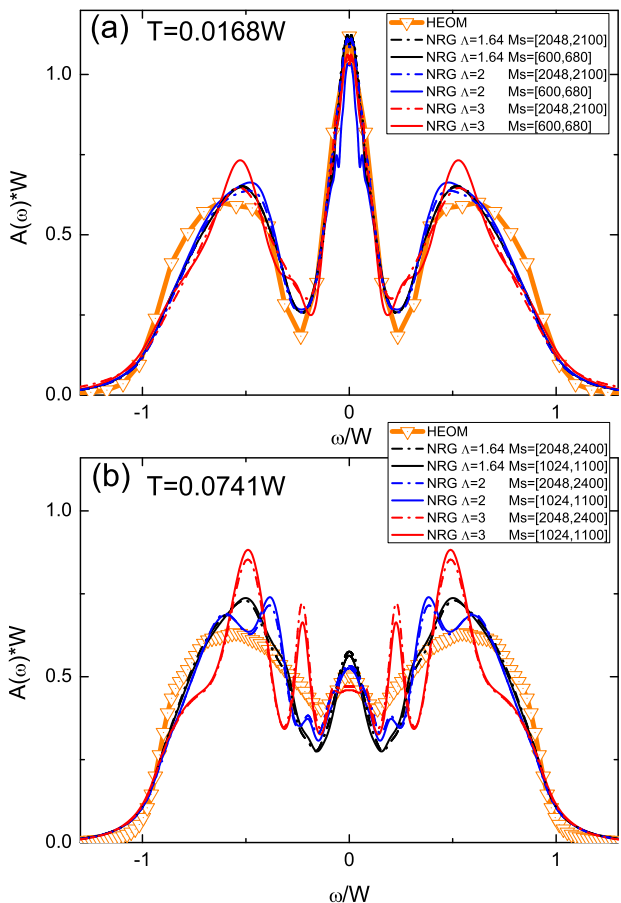}
\caption{(Color online). Spectral functions obtained by the NRG approach used in Ref.~\onlinecite{Bul01045103} with different values of
the controlling parameters $\Lambda$ and $M_s$, at $T=$ (a) $0.0168W$ and (b) $0.0741W$.
The expression $M_s=[a,b]$ with numbers $b>a$ means that the actual value of $M_s$ is chosen in the interval $[a,b]$ with the largest energy gap.
The hybridization functions are extracted from the HEOM converged results, and are used for both impurity solvers.
The corresponding HEOM calculated spectral functions are also shown for comparison. } \label{fig11}
\end{figure}

For the HEOM approach, there are only two controlling parameters, the memory component $M$ in Eq.~(\ref{def-M}) and the truncation level $L$.
Its convergence with respect to these two parameters have already been affirmed; see Figs.~\ref{fig2},~\ref{fig3} and~\ref{fig4}.
It can be seen that the convergence of the HEOM approach is much easier at high temperature than that at low temperature,
as by construction it is intrinsically more favorable at higher temperature cases.
The convergence of all the HEOM results shown in \Fig{fig10} and \Fig{fig11} have also been carefully checked.

In contrast, the NRG approach used in Ref.~\onlinecite{Bul01045103} involves several controlling parameters.
The most important two are the discretization parameter $\Lambda$ and the number of kept states $M_s$.
The NRG results are considered to be numerically exact if they converge as $\Lambda\rightarrow1$ and $M_s\rightarrow\infty$.
However, it is usually very difficult to reach at high temperature,
due to the exponential increase of the computational cost.
The NRG data\cite{Bul01045103} we cited in Fig.~\ref{fig10} were obtained under $\Lambda=1.64$,
and the convergence was not reported.
Therefore, we carry out a check of its convergence for both low and high temperatures; see Fig.~\ref{fig11}.

Figure~\ref{fig11} shows that the NRG results converge rapidly
with respect to $\Lambda$ and $M_s$ at low temperature, but rather slowly at high temperature,
which is very different (or opposite to) the HEOM approach.
At the low temperature $T=0.0168W$, the overall shape of the best NRG curve with $\Lambda=1.64$ and $M_s=[2048,2100]$
is very close to the HEOM converged spectral function, which is consistent with the fact that at low temperature the data in Fig.~\ref{fig10} are almost identical.
Here $M_s=[2048,2100]$ means that the actual value of $M_s$ is chosen between 2048 and 2100 with the largest energy gap.
At high temperature $T=0.0741W$, the best NRG spectral function with $\Lambda=1.64$ and $M_s=[2048,2400]$ in our calculation is still not guaranteed to be converged.
Moreover, its overall shape dramatically differs from the converged HEOM one.
It indicates that large uncertainties reside in the $T=0.0741W$ NRG data shown in Fig.~\ref{fig10}.

It should be pointed out that the NRG data of Ref.~\onlinecite{Bul01045103}
used for the comparison in \Fig{fig10} do not represent the best possible NRG results.
In the past decade, the quality of the spectral function produced by NRG
has been improved dramatically by the new algorithms
such as the density matrix NRG\cite{Hof001508}
and the full density matrix NRG.\cite{Wei07076402,Pet06245114}
We have also checked the convergence of the corresponding spectral function at $T=0.0741W$
using the full density matrix NRG approach; see Supplemental Material.\cite{Sup}
The resulted spectral function is overall much closer to the HEOM counterpart,
including the heights of the Kondo and the Hubbard peaks,
and the overall line shape of the dip between them.\cite{Sup}
%

%\textbf{
At this stage, the origin of the remaining difference between the HEOM and NRG spectral functions displayed in \Fig{fig11}(a) is unclear.
Clarifying this issue requires a more careful and comprehensive assessment of the NRG methods (since the HEOM data have converged). However, this could be technically very difficult, and is beyond the scope of this paper.
Therefore, this problem is left open for further investigations.
%}

\section{Concluding Remarks}\label{sec4}

To summarize, the HEOM approach is used as the impurity solver for DMFT method.
This method is employed to investigate the Mott-Hubbard metal-insulator transition
in both the hypercubic and the Bethe lattices at finite temperature.
%
%\textbf{
At low temperatures the results obtained by the HEOM based DMFT method agree well with those using the NRG method as the impurity solver.
At higher temperatures, the HEOM approach becomes much more efficient. Thus the HEOM approach provides a complement to the NRG impurity solver,
in the sense that it is highly efficient and converges rapidly at finite temperatures.
%}

Currently the developed approach is limited to a temperature no lower than $T=0.0125W$.
The reason is that at lower temperature the numeric convergence requires larger
truncation level.
It is however possible to design more efficient reservoir memory decomposition schemes
to dramatically reduce the computational resources requirements.
The Lorentzian fit scheme for evaluating the hybridization function may also
introduce minor residual error under certain circumstances.
For example, the quasi-particle peak at the Fermi level becomes quite narrow
when the system is very near to the metal-insulator transition point.
To distinguish this type of peak in the process of Lorentzian fit,
much finer frequency meshes near the Fermi level is necessary.
It will sometimes lead to minor fitting error for the narrow quasi-particle peak.

The HEOM approach adopts a general form of the system Hamiltonian.
It is quite convenient to extend the current HEOM+DMFT approach
to systems other than the half-filled single-band situation.
For example, the system occupation number can be tuned away from half-filling
by the adjustment of the chemical potential, to simulate the influence of
charge doping to SCS.
In this situation, the asymmetric Anderson impurity model is to be employed in
the HEOM impurity solver,
and an additional search is required for the correct chemical potential
that reproduces the true occupation number on a lattice site.

The HEOM approach is also applicable to complicated multi-orbital impurity models.
For instance, the HEOM approach has been applied to a three-impurity Anderson model;
see Fig.~S8 in the Supplemental Material of Ref.~\onlinecite{Li12266403}.
This confirms the potential applicability of the HEOM method to more complex systems.
Moreover, it is possible to improve further the efficiency of HEOM approach
based on physical considerations and making use of the sparse nature of HEOM.
This may extend further the applicability of HEOM+DMFT method
to further complex strongly correlated systems.

\acknowledgments
The support from the National Science Foundation of China (Grants No.\,21303175, No.\,21103157,
No.\,21033008, No.\,21233007, No.\,11074302, No.\,11374363, and No.\,21322305),
%the Fundamental Research Funds for Central Universities of China (Grants No.\,2340000034 and No.\,2340000025),
the Strategic Priority Research Program (B) of the CAS (Grant No.\,XDB01020000),
the National Program on Key Basic Research Project of China (Grant No.\,2012CB921704),
and the Hong Kong UGC (Grant No.\,AoE/P-04/08-2) and RGC (Grant No.\,605012)
is gratefully appreciated.

%\iffalse
%\begin{thebibliography}{10}
%\end{thebibliography}
%\fi

%\bibliographystyle{aip}
%\bibliography{refs-DMFT}

\end{document}